\documentclass[aps,prb,reprint,superscriptaddress,showkeys]{revtex4-1}

\usepackage{color}
\usepackage{graphicx}
\usepackage{dcolumn}
\usepackage{bm}
\usepackage{verbatim}
\usepackage{upgreek}

\begin{document}


\title{Imaging Flux Vortices in MgB$_{\text{2}}$ using Transmission Electron Microscopy}


\author{J.C. Loudon}

\author{C.J. Bowell}
\altaffiliation[Now at ]{Halcyon Molecular, 505 Penobscot Drive, Redwood City, CA 94063, USA.}

\affiliation{Department of Materials Science and Metallurgy, University of Cambridge, Pembroke Street, Cambridge CB2 3QZ, United Kingdom}

\author{N.D. Zhigadlo}

\author{J. Karpinski}
\affiliation{Laboratory for Solid State Physics, ETH Zurich, Schafmattstrasse 16, CH-8093, Zurich, Switzerland}

\author{P.A. Midgley}
\affiliation{Department of Materials Science and Metallurgy, University of Cambridge, Pembroke Street, Cambridge CB2 3QZ, United Kingdom}

\date{\today}

\begin{abstract}

We report the successful imaging of flux vortices in single crystal
MgB$_{\text{2}}$ using transmission electron microscopy. The specimen
was thinned to electron transparency (350~nm thickness) by focussed
ion beam milling. An artefact of the thinning process was the
production of longitudinal thickness undulations of height 1--2~nm in
the sample which acted as pinning sites due to the energy required for
the vortices to cross them. These had a profound effect on the
patterns of vortex order observed which we examine here.
\end{abstract}


\maketitle


\section{Introduction}

Superconductors have zero electrical resistance and expel magnetic
flux from their interiors (the Meissner effect). If a magnetic field
is applied to an ideal (type I) superconductor, no flux enters unless
the field exceeds the critical field, $H_c$, at which the material
ceases to be superconducting. However, in type II superconductors, the
whole superconducting state is not destroyed at once but above the
lower critical field, $H_{c1}$, magnetic flux penetrates the
superconductor by flowing along channels called flux vortices. The
vortices consist of a core where superconductivity is suppressed with
a size given by the coherence length, $\xi$, surrounded by circulating
supercurrents which persist over a distance called the penetration
depth, $\Lambda$. The magnetic flux associated with the vortex
persists over the same distance. Each vortex carries a single quantum
of magnetic flux given by $\Phi_0=h/2e$ where $h$ is Planck's constant
and $e$ is the electron charge. Vortices repel one another and in a
defect-free, isotropic superconductor will form a 2-dimensional
hexagonal lattice called an Abrikosov lattice \cite{Abrik57}.

MgB$_2$ is a superconductor discovered in 2001 \cite{Nag01} with a
transition temperature $T_c=39$~K. It has a hexagonal crystal
structure (space group 191: $P6/mmm$) with $a=b=3.086$~{\AA} and
$c=3.542$~{\AA} with alternating layers of magnesium and boron. There
is disagreement in the literature on the values of $\Lambda$ and $\xi$
at zero temperature and field with $\xi_{ab}$ ranging from 5--13~nm
and $\Lambda_{ab}$ ranging from 48--130~nm \cite{Mosh09, Cubitt03,
  Manz02}. The values in the $c$ direction are less well reported but
Moshchalkov {\it et al.}  \cite{Mosh09} give $\xi_c=51$~nm and
$\Lambda_c=33.6$~nm. It is an unusual superconductor as it has two
types of electronic bonding: $\sigma$-bonding from the boron $p_{xy}$
orbitals and $\pi$-bonding from the boron $p_z$ orbitals
\cite{Cubitt03}, which give rise to two superconducting gaps the
magnitudes of which vary in the literature with $\Delta_\pi=1.2$--3.7~meV
and $\Delta_\sigma=6.4$--7.2~meV \cite{Brandt11}.

\begin{figure}
\includegraphics[height=55mm]{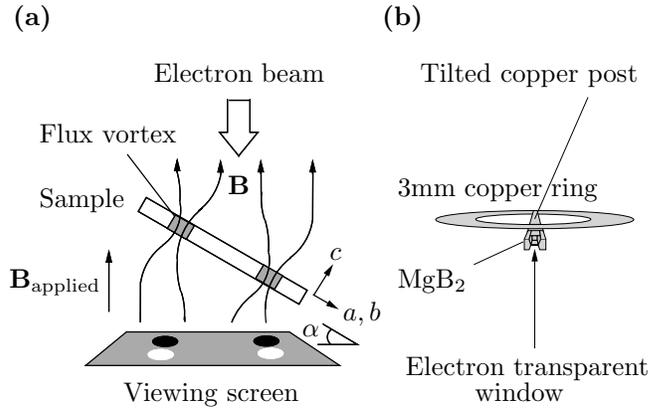}
\caption{\label{fig1} (a) Experimental arrangement for imaging flux
  vortices. The electrons are deflected by the component of the
  B-field from the vortices normal to the electron beam giving a
  black-white feature in an out-of-focus image. (b) The specimen
  geometry. The MgB$_2$ specimen was mounted to a copper post glued to a
  standard 3~mm diameter copper ring at an angle of $45^{\circ}$. An
  electron transparent window was then cut by focussed ion-beam
  milling.}
\end{figure}

Flux vortices can be imaged using transmission electron microscopy
\cite{Har92} due to the magnetic field from the vortices deflecting
the electron beam and appear as black-white features in an
out-of-focus image (see Fig.~\ref{fig1}(a)). This is a unique imaging
technique as it gives information on the internal structure of the
vortices, not just the stray fields, and allows imaging at video
rate. In this paper, we study the ordering of flux vortices in MgB$_2$
using this technique and will report on the internal magnetic
structure of the vortices and their dynamics in subsequent
publications.

\section{Sample Preparation and Experimental Methods}

MgB$_2$ single crystals were synthesised by Dr J. Karpinski as
described in ref.~\cite{Karp03} and were thinned to 250~nm in the $c$
direction using focussed ion-beam (FIB) milling so that they were
electron transparent. The sample needs to be tilted to a high angle
($\alpha$ in Fig.~\ref{fig1}(a)) to maximise the electron beam
deflection and so was attached to a tilted copper post as illustrated
in Fig.~\ref{fig1}(b) giving a tilt angle of $\alpha=45 \pm 1^\circ$
and a projected thickness of 350~nm parallel to the electron beam. A
liquid-helium cooled specimen stage was used to cool the sample and
energy-filtered images were recorded with a CCD camera using a Philips
CM300 transmission electron microscope equipped with a field-emission
gun operated at 300~kV. A magnetic field was applied to the sample by
altering the setting of the twin lens in the microscope. Further
details are given in supp. info. 1.

\section{Results and Discussion}

\begin{figure*}

\includegraphics{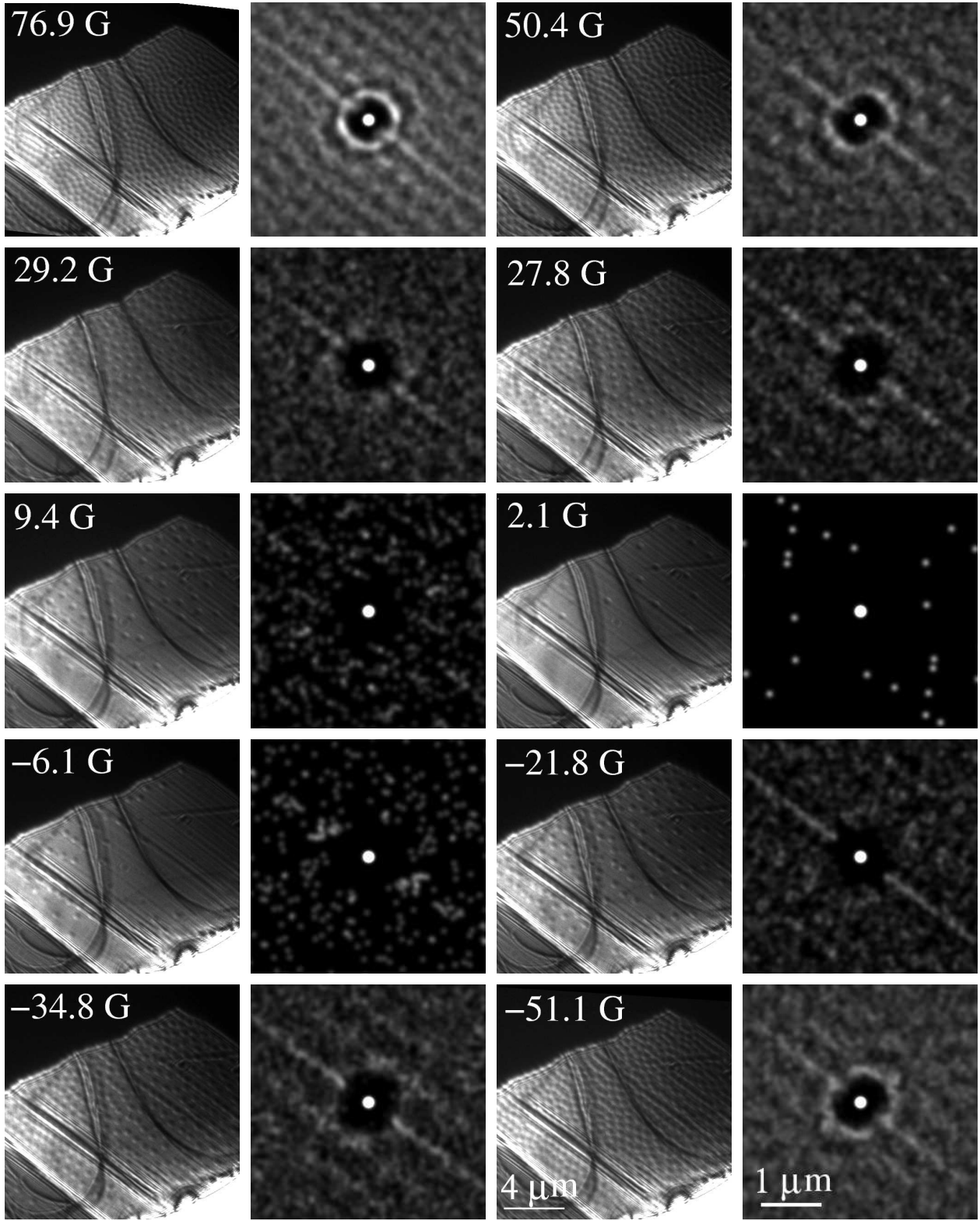}

\caption{\label{fig2} Sequence of tilt-compensated images showing flux
  vortices in MgB$_2$ taken at 10.8~K with a defocus of $3.46 \pm
  0.02$~cm at different B-fields (indicated on the images). The edge
  of the specimen is at the bottom-right of the image with vacuum
  beyond. The black area at the top-left is the bulk MgB${_2}$ crystal
  to which the electron transparent window is attached. The
  autocorrelation function of the vortex positions is shown to the
  right of each image.}

\end{figure*}

Fig.~\ref{fig2} shows images of flux vortices in MgB$_2$ taken at
10.8~K in different magnetic fields. The B-fields quoted are
calculated from the vortex density averaged across the
electron-transparent window. Energy-filtered imaging was used to
produce a thickness map of the specimen (see supp. info. 1) showing
that the specimen had a thickness parallel to the electron beam of
100~nm at the edge nearest the vacuum which gradually increased to
370~nm in the centre and then fell back to 330~nm at the top-left of the
image due to a slight undercutting during FIB milling. The images have
been `tilt-compensated' to correct for the apparent compression of the
vortex spacings due to the specimen tilt as described in
ref.~\cite{Lou09} and the originals are in supp. info. 2.

An unintended effect of the FIB milling was to produce longitudinal
thickness undulations running parallel to the ion beam which appear as
lines running from top-left to bottom-right of the images in
Fig.~\ref{fig2}. The thickness map showed that the undulations were
1--2~nm in height and that the dark strip near the bottom-left was
20~nm in height. Interestingly, these act as pinning sites as a vortex
needs to become longer to move onto a thicker part of the undulation.

In supp. info. 3 we estimate that it requires $\sim 0.5$~eV to
increase the length of a vortex line by 1~nm in MgB${_2}$ and that the
energy cost in moving a vortex from its equilibrium position in an
ideal Abrikosov lattice is of a similar magnitude for the
displacements we see here indicating that the vortex arrangement is a
competition between these two energies. In contrast, the thermal
energy of the vortices, $k_BT=0.0009$~eV, is much lower.

The sequence in Fig.~\ref{fig2} begins at 76.9~G and the field is
reduced in subsequent images. Antivortices are seen when the field
goes negative as indicated by the reversal of the black-white
contrast. To assess the ordering of the vortices, autocorrelation
functions of the vortex positions, found by motif
matching\cite{Lou09}, are shown to the right of each image. Histograms
showing the distribution of vortex spacings and the angle which lines
connecting neighbouring vortices (`bonds') make to the horizontal are
shown in supp. info. 4.


When the field is taken through zero and goes negative, the
antivortices first appear, not at the free edge as might be expected,
but furthest away from this where the electron-transparent window
meets the bulk sample. As the magnitude of the field is increased,
they travel in lines down certain thickness undulations to reach the
rest of the sample. The phenomenon of a vortex-free region near the
sample edge has been noted by Olsen {\it et al.}~\cite{Olsen04} who
use magneto-optical imaging to visualise vortices in NbSe$_2$ and find
that a vortex-free region of width 5~$\upmu$m is seen at the sample
edge in an applied field of 2~Oe. The vortex-free region is much
larger than the penetration depth (265~nm in NbSe$_2$)
and is ascribed to the vortices forming at the sample edge and then
immediately being pushed into the specimen interior by the Lorentz
force from the B-field which penetrates the superconductor near the
edge~\cite{Olsen04, March95}. As the applied field is increased, new
vortices entering the sample are repelled by those which have aleady
accumulated in the interior and so the vortex-free region diminishes
as the field is increased as observed both here and in
ref.~\cite{Olsen04}. We are currently undertaking dynamic experiments
to clarify how exactly the vortices first enter the sample.


All the autocorrelation functions show a streak at $142^{\circ}$
anticlockwise from the horizontal indicating a preference for the
vortices to align along the thicknes undulations. This `vortex
channelling' has also seen using scanning SQUID microscopy for
thickness trenches of depth 25--125~nm in amorphous MoGe~\cite{Plourde02}.

The image for -6.1~G shows strikingly different spacings parallel and
perpendicular to the undulations with the histograms in supp. info. 4
showing a peak for spacings between 856--963~nm corresponding to the
vortex spacing parallel to the length of the undulation and peaks at
1070--1177~nm and 1284--1391~nm for spacings perpendicular to this. As
the field is increased, the vortex spacings parallel and perpendicular
to the undulations become the same as shown by the autocorrelation
functions in Fig.~\ref{fig2} becoming circular and the histograms of
vortex spacings in supp. info. 4 becoming single peaked.

There is an absence of regularly-spaced peaks in the autocorrelation
functions in Fig.~\ref{fig2} indicating that the vortices do not form
an ordered lattice. However, the autocorrelation for 76.9~G shows weak
peaks which occur at angles of $10^\circ$, $55^\circ$, $107^\circ$ and
$143^\circ$ in its inner ring (Supp. Info. 5). These are approximately
$45^\circ$ apart, indicating a tendency towards a square lattice with
two different `grains' oriented at $45^\circ$ to one another rather
than the hexagonal Abrikosov lattice expected in the absence of
pinning.  The histogram of bond directions in Supp. Info. 4 indicates
there is a similar fraction of each grain type. Grain boundaries can
be seen between different types of vortex order in the image:
sometimes these resemble sharp stacking faults but there are also arcs
of vortices which give a gradual change in orientation. The order
within each grain does not seem to be very good as an exponential
decay fit to the peaks of correlation function yields a correlation
length of 1.1 vortex spacings.

\section{Conclusions}

We have successfully imaged flux vortices in single crystal MgB$_2$
using transmission electron microscopy. The ion-thinning process used
to thin the sample to electron transparency (350~nm thickness)
produced longitudinal thickness undualtions 1--2~nm in height creating
a pinning landscape which profoundly affected the movement of the
vortices and the arrangements they adopted. As the applied magnetic
field was increased from zero, flux vortices first appeared in the
part of the electron transparent window furthest from the specimen
edge. This is probably due to the vortices forming on the specimen
edge and then being pushed into the interior by the Lorentz force from
the B-field which penetrates near the edge as described in
ref.~\cite{Olsen04}. As the field was further increased, the vortices
moved along the length of the thickness undulations to reach the rest
of the sample. The vortices tended to align in rows along the
thickness undulations but no well-ordered lattice was observed at the
B-fields used here (0--76.9~G). However, at the highest field of
76.9~G, autocorrelation functions of the vortex positions showed a
tendency for the vortices to form a square lattice rather than the
hexagonal Abrikosov lattice expected in the absence of pinning.

\begin{acknowledgments}
This work was funded by the Royal Society and the EPSRC, grant number
EP/E027903/1.
\end{acknowledgments}




\providecommand{\noopsort}[1]{}\providecommand{\singleletter}[1]{#1}%

\end{document}